\documentstyle[psfig,aps,prl,amssymb]{revtex}

\begin{document}
\title{Increased Efficiency of Quantum State Estimation Using 
 {\it Non-Separable} Measurements}
\author{Paul B. Slater}
\address{ISBER, University of California, Santa Barbara, CA 93106-2150\\
e-mail: slater@itp.ucsb.edu, FAX: (805) 893-7995}

\date{\today}

\draft

\maketitle

\vskip -0.1cm

\begin{abstract}
We address the ``major open problem'' of evaluating how much increased
efficiency in estimation is possible using {\it non-separable} --- as 
opposed to separable --- measurements of $N$ copies 
of $m$-level quantum systems. First, we study the six cases 
$m=2$, $N=2,\ldots,7$ by 
computing  the $3 \times 3$ Fisher information
matrices for the corresponding {\it optimal} measurements 
recently devised  by Vidal {\it et al}
(Phys. Rev. A 60, 126 [1999]). We obtain simple polynomial expressions for 
the (``Gill-Massar'')  traces of the 
products of the inverse of the 
quantum Helstrom information matrix and these
Fisher information matrices.
The six traces {\it all}
have {\it minima} of $2 N -1$
in the {\it pure state} limit --- while for {\it separable} measurements 
(Phys. Rev. A 61, 042312 [2000]), 
the traces can equal $N$, but {\it not} exceed it. 
Then, the result of an analysis for $m=3$, $N=2$ 
leads us to {\it conjecture} that for 
optimal measurements for {\it all}
$m$ and $N$, the Gill-Massar trace achieves a 
{\it minimum} of  
$(2 N -1) (m-1)$ in the {\it pure state} limit.
\end{abstract}


\vspace{.1cm}

\tableofcontents

\section{Introduction}
We investigate information-theoretic properties of the optimal
measurement schemes recently devised by Vidal {\it et al} 
\cite{vidal}, helping thereby to address the ``major open problem'' 
\cite{gill} of evaluating how much 
increased efficiency in estimation 
is possible using {\it non-separable} measurements (cf. \cite{fischer}). 
In their 
extensive study, ``State estimation for large ensembles,'' 
which we seek to extend here, Gill and
Massar stated that ``we cannot compare our results with the recent analysis
of covariant [optimal] measurements on mixed states \cite{vidal} 
because we suppose separability
of the measurement, whereas \cite{vidal} does not'' \cite{gill}.
A ``separable measurement is one that can be carried out sequentially
on separate particles, where the measurement on one particle at any stage
(and indeed which particle to measure: one is allowed to measure particles
several times) can depend arbitrarily on the outcomes so far'' \cite{gill}.

The analyses  here are conducted in terms of the (classical) {\it Fisher 
information} (of the probability distributions associated with
the non-separable measurements), making use of 
the quantum (Helstrom) Cram\'er-Rao bound 
\cite{helstrom} on the Fisher information matrix 
for any {\it oprom} (operator-valued probability measure)
\cite{gill2,busch}.
Contrastingly, the studies of Vidal and his several Barcelona colleagues 
\cite{vidal,tarvid,vidal2,acin}
 have been formulated primarily in terms of
{\it fidelity}, $F(\rho,\rho')$
($\rho$ and $\rho'$  being density matrices) \cite{uhlmann,jozsa},  
and secondarily, {\it information gain} \cite{tarvid}.
Now, there surely exists
an intimate connection between these approaches, since 
$2(1-F(\rho,\rho'))$ functions as the {\it Bures} distance between
$\rho$ and $\rho'$. The Bures metric is a distinguished member (the {\it 
minimal}
one) of a continuum of possible quantum extensions --- each 
associated with a distinct {\it operator monotone} function --- of the 
(classical) Fisher information
metric \cite{petzsudar,bc,paulpla}. The Helstrom-Cram\'er-Rao bound
corresponds to the particular use of the Bures metric {\it via} the concept of 
the {\it symmetric logarithmic derivative} \cite{helstrom}.
An interesting hypothesis is that asymptotically the Fisher information
matrix for optimal measurements is simply proportional to the metric
tensor associated with some specific operator monotone function.
(Our results below indicate that such a role is 
definitely {\it not} played by the
Bures metric.)

We shall be concerned 
here primarily (cf. secs.~\ref{tl} and \ref{fl}) with the
two-level quantum systems, representable by the $2 \times 2$ density matrices,
\begin{equation} \label{bloch}
\rho  = {1 \over 2} \pmatrix{1 + z & x + \mbox{i} y \cr
x - \mbox{i} y & 1- z \cr},
\end{equation}
where $r^2 = 
x^2 +y^2 +z^2 \leq 1$. The particular $(x,y,z)$ parameterization employed
in (\ref{bloch})
 corresponds to  the use of Cartesian coordinates for the  ``Bloch 
(or Poincar\'e) sphere'' (unit ball in 
three-space) representation of the two-level systems \cite{bm} 
\cite[sec. 4.2]{belt}, while the alternative (spherical coordinate)
parameter
$r$ is  the radial distance from the origin. Pure states, for which 
$|\rho|=0$, correspond to
$r=1$ and the fully mixed state, for which $|\rho| = {1 \over 4}$, to $r=0$.

For the cases of $N$  copies ($N=2,\ldots,7$) of a two-level quantum system
(\ref{bloch}) we obtain below in sec.~\ref{relations} 
a quite interesting pattern of results of increased 
efficiency using non-separable measurements,
which strongly suggests generalizability to arbitrary $N$.
To explicitly examine the cases $N>7$
would either entail considerable additional computations
for each specific $N$ and/or substantial analytical advances 
(cf. sec.~\ref{fishmono}) allowing one
to formally establish the measure of increased efficiency 
for {\it arbitrary} $N$. (We note that Latorre {\it et al} \cite{vidal2}
had to proceed {\it case-by-case}, 
that is, each $N$ individually, since they ``did not know how to build the
POVM algorithmically''.) In sec.~\ref{fishmono} we explore one possible
approach in this regard, attempting to explain the Fisher information 
matrices we compute in sec.~\ref{nce} in terms of monotone metrics. 
In sec.~\ref{nce}, we also formulate a conjecture as to the increase in
efficiency achieveable using non-separable optimal 
measurements for $N$ copies of $m$-level
quantum systems in general.

To begin our study, immediately below in sec.~\ref{goforit}, 
we  expand upon an observation \cite[p. 2684]{slatjmp} regarding 
 an information-theoretic relationship 
between certain classical
and quantum entities --- that is, the Fisher information matrix for 
a certain (quadrinomial) multinomial probability distribution and 
the quantum Helstrom information matrix (proportional to the 
Bures metric tensor), and its implications for 
optimal measurements.

 In sec.~\ref{uc} we examine further 
ramifications on issues 
of state estimation \cite{gill,helstrom} and 
 universal coding (data compression) \cite{cb1,kratt,kratt2,jozsa2}.
There appears to be an interesting relation between the devising of
optimal measurements as in \cite{vidal}, 
and universal quantum coding, as both processes involve
averaging with respect to isotropic prior probability distributions 
by ``projecting onto total spin eigenspaces, and within each such subspace,
onto total spin eigenstates with maximal total spin component in some 
direction'' \cite{vidal} --- cf. \cite[eqs. (5.33) and (5.34)]{vidal} 
and \cite[eq. (2.48)]{kratt}. The particular prior distribution which 
yields both the minimax and maximin for the universal quantum coding of
the two-level systems is based on the {\it quasi-Bures} metric, a particular
example of a monotone metric. We attempt in sec.~\ref{fishmono} 
to relate the Fisher information 
matrices we compute in sec.~\ref{nce} to the monotone metrics.
\section{Proportionality between Helstrom and Fisher Information 
Matrices} \label{goforit}
The density matrices (\ref{bloch})
turn out to have an intimate relationship
with a particular form of multinomial (that is, 
quadrinomial) probability distributions --- the 
{\it four} distinct possible outcomes
being  assigned probabilities
\begin{equation} \label{qpd}
 x^2,\quad y^2,\quad z^2, \quad 1-x^2-y^2-z^2 .
\end{equation}
One can attach to the three-dimensional convex set of two-level 
quantum systems (\ref{bloch}), 
adapting  one (the simplest) of the ``explicit'' 
formulas of Dittmann \cite[eq. (3.7)]{ditt1} \cite{ditt2},
\begin{equation}
d_{Bures}(\rho,\rho + \mbox{d} \rho)^2 =
{1 \over 4} \mbox{Tr} \{ \mbox{d} \rho \mbox{d} \rho +{1 \over |\rho|}
(\mbox{d} \rho  - \rho \mbox{d} \rho ) (\mbox{d} \rho -\rho \mbox{d} \rho) \},
\end{equation}
the $3 \times 3$ quantum (Helstrom) information
matrix \cite{helstrom,gill,barn}
 (that is, {\it four} times 
the Bures metric tensor \cite{ditt2,hub1,hub2,bc}),
\begin{equation} \label{niu}
H_{q}(x,y,z) = {1 \over  (1-x^2-y^2-z^2)} \pmatrix{1-y^2-z^2 & x y & x z \cr
x y & 1- x^2 -z^2 & y z \cr
x z & y z & 1-x^2 -y^2 \cr}.
\end{equation}
We  use the subscripts $q$ and $c$ --- in a suggestive, perhaps not
fully rigorous manner --- to denote results stemming from quantum or
classical considerations. Also, note that (\ref{niu}) ``blows up'' at the
pure states themselves --- so it will be problematical, at best, to 
directly compare
results pertaining to (\ref{niu}) with ones based on {\it pure state}
models \cite{gill,fuji}.

In spherical coordinates $(r,\theta,\phi$), $x = r \cos{\theta},
y =r \sin{\theta} \cos{\phi}, z = r \sin{\theta} \sin{\phi}$, the
matrix (\ref{niu})  takes a
 {\it diagonal} form,
\begin{equation} \label{sPh}
H_{q}(r,\theta,\phi)  =  \pmatrix{ {1 \over 1 - r^2} & 0 & 0 \cr
0 & r^2 & 0 \cr
0 & 0 & r^2 \sin^2{\theta} \cr},
\end{equation}
for this {\it orthogonal} system of coordinates (cf. \cite{tod}).
(Below, in the interest of succinctness, 
we will replace the frequently-occurring 
expression $x^2+y^2+z^2$ by its equivalent, $r^2$.)

Now,  the quantum information matrices (\ref{niu}) and (\ref{sPh})
are   simply proportional to the (classical) Fisher information 
\cite{frieden} matrices $I_{c}(x,y,z)$ and $I_{c}(r,\theta,\phi)$
for the quadrinomial probability distribution (\ref{qpd}).
(By way of algorithmic example, the $xy$-entry  of the $3 \times 3$
Fisher information matrix --- in its Cartesian coordinate form, 
$I_{c}(x,y,z)$ --- is
 computable as the expected value
of the 
[two-fold] product of the logarithmic derivatives of (\ref{qpd})
 with respect to
$x$ and with respect to $y$.)
More precisely, the nine entries of  $I_{c}(x,y,z)$  are
all {\it four} times the 
corresponding entries of (\ref{niu}), that is
\begin{equation}
I_{c}(x,y,z) = 4 H_{q}(x,y,z).
\end{equation}
A natural explanation for this  phenomenon is that the 
{\it information geometry} \cite{murray}
 of both models is that of the standard metric on
the surface of a three-sphere in four-dimensional Euclidean space 
\cite{bc,kass}.

Both quantum (Helstrom) information and Fisher information 
possess the property of {\it additivity}, that is, for $N$ independent 
identical density matrices
or probability distributions, the information matrices 
(possibly scalars) are $N$ times those
for a single one  \cite[exer. 1.10]{gill2}
\cite[sec. VI.4]{helstrom}
 \cite{kagan,chentsov,kagan2,rao}.

By the quantum version of the Cram\'er-Rao theorem \cite{helstrom},
the inverse matrix 
$H_{q}(x,y,z)^{-1}$  serves as a lower
bound on the variance-covariance matrix $V(x,y,z)$
for any {\it unbiased}
estimator of the parameters ($x,y,z$) of $\rho$.
(This means that the matrix 
difference, $V(x,y,z) -H_{q}(x,y,z)^{-1}$, must be  nonnegative definite,
that is, have all its
eigenvalues nonnegative.)
 In this regard,
\begin{equation} \label{inv}
H_{q}(x,y,z)^{-1} =  \pmatrix{1 -x^2 & -x y & - x z \cr
- x y & 1-y^2 & - y z \cr
-x z & - y z & 1-z^2 \cr}
\end{equation} 
(Of course,  $H_{q}(r,\theta,\phi)^{-1}$ is diagonal.)

By dint of the  additivity of information, in conjunction with the
 Cram\'er-Rao theorem (cf. \cite[eq. (26)]{gill}), one can 
conclude that it is {\it not}
 possible to devise
for  $N < 4$ independent identical two-level systems, an {\it oprom}
\cite{gill2,busch}, which has
for its outcomes the quadrinomial distribution (\ref{qpd}) 
(cf. \cite{vidal,bennett}).
(When we attempted to construct such an oprom for the 
case $N=2$, we found that the four
 operators could {\it not} all be nonnegative 
definite if they were to yield (\ref{qpd}).) However, for
$N \geq 4$, the question 
of whether such an oprom exists would appear
 to be a completely open one --- since now
the Cram\'er-Rao theorem does {\it not} rule out its possibility.
(The results of Vidal {\it et al} \cite{vidal}
show that an optimal {\it minimal} number of measurements for $N>3 $ is
at least {\it fifteen},
 exceeding  the number {\it four} for  an oprom that would give 
as its outcomes, the quadrinomial probability 
distribution (\ref{qpd}).) If such an oprom could be found for $N=4$ itself,
then the Cram\'er-Rao inequality would be {\it fully} saturated.
\section{Analyses of {\it Optimal} Measurements of Vidal {\it et al} for
$N$ Copies of Two-Level Quantum Systems} 
\label{nce}
\subsection{Computation of the Fisher Information Matrices} \label{omer}
\subsubsection{$N=2$}
Let us now consider the probability distribution
 in \cite{vidal} obtained from the optimal minimal number (five) of
measurements for the case of $N=2$ identical independent copies of
the two-level systems (\ref{bloch}). 
The  five probabilities --- as we have explicitly found --- 
can be written as (the three)
\begin{equation} \label{ped}
{1 \over 4} (1-r^2),\quad  {3 \over 16} (1+z)^2, \quad {1 \over 48}
(8 x^2 -4 \sqrt{2} x (z-3) +(z-3)^2) ,
\end{equation}
together with the pair
\begin{displaymath}
{1 \over 48} (9 + 2 x^2 \pm 4 \sqrt{3} x y + 6 y^2 + 2 \sqrt{2}
(x \pm \sqrt{3} y) (z-3) -6 z + z^2).
\end{displaymath}

Quite remarkably, the associated Fisher information matrix 
($\tilde{I}_{c}$) turns out
to  precisely equal the quantum (Helstrom) information matrix,
$H_{q}(x,y,z)$  --- and not $2 H_{q}(x,y,z)$, which is the
 upper bound furnished by the
quantum Cram\'er-Rao theorem. So, the bound could be said to be
``half-saturated''.
(In regard to this specific result, R. Gill has observed that there may
exist other measurement schemes which are {\it sub-optimal} accoding to the 
{\it fidelity} criterion of \cite{vidal}, but superior
 in terms of Fisher information (cf. \cite{tarvid}).)
\subsubsection{$N=3$}
For an optimal minimal set of measurements for $N=3$, we can take the eight
probabilities, consisting of the four pairs,
\begin{equation}
 {(1 \pm x)^3 \over 12},\quad
{(1 \pm y)^3 \over 12}, \quad {(1 \pm z)^3 \over 12}, \quad {1 \over 4} 
(1 \pm {x+y+z \over \sqrt{3}})  (1 -  r^2) .
\end{equation}
The associated Fisher information matrix is expressible as
\begin{equation} \label{pkd}
2 H_{q}(x,y,z) + {1 \over 2( (x+y+z)^2 -3)}  \pmatrix{a & b & b \cr
b & a & b \cr
b & b & a \cr},
\end{equation}
where $a = 2 (1-x y - x z - y z)$ and $b=-1+r^2$.
The second summand in (\ref{pkd}) is {\it negative} definite (having two 
of its three negative
eigenvalues equal to $-{1 \over 2}$), while $3 H_{q}(x,y,z)$ is the upper bound
on the Fisher information matrix 
provided by the Cram\'er-Rao theorem.
\subsubsection{$N=4$}
An optimal minimal set of measurements for $N=4$ yields a 
fifteen-vector of probabilities. The Fisher information matrix for this
probability distribution is
\begin{equation} \label{cue}
3 H_{q}(x,y,z) + {1 \over 12} \pmatrix{-7-5 y^2 - 5 z^2 & 5 x y & 5 x z \cr
5 x y & -7 - 5 x^2 - 5 z^2 & 5 y z \cr
5 x z & 5 y z & -7 -5 x^2 - 5 y^2 \cr}.
\end{equation}
The second term is {\it negative} definite with one eigenvalue
equal to $-{7 \over 12}$ and the other two, $ -{1  \over 12} 
(7 + 5 r^2)$. If we subtract (\ref{cue}) from 
the Cram\'er-Rao upper bound $4 H_{q}(x,y,z)$, we obtain (as we must) 
a nonnegative definite
matrix, having  two eigenvalues
 ${1 \over 12} (19 + 5 r^2)$ and one,
${7 \over 12} + {1 \over 1 - r^2}$.
\subsubsection{$N=5$}
For $N=5$, a twenty-vector of probabilities was obtained for the optimal 
minimal number of measurements. The Fisher information matrix can be 
expressed as the sum of $4 H_{q}(x,y,z)$ (which dominates it, while
$3 H_{q}(x,y,z)$ does not) 
and a {\it negative}
definite matrix, having one of its three negative eigenvalues equal to
$-{3 \over 16} (5+3 r^2)$. This negative definite matrix
can be written as the product of ${1 \over 16 (-3 +(x+y+z)^2)}$ 
and a $3 \times 3$  matrix,
the $(1,1)$  cell of which is
\begin{equation} \label{fds}
-2 (-20 + 7 y^4 +9 y^3 z - 11 z^2 + 7 z^4 - 5 x^3 (y+z) + 3 y z (5 + 3 z^2) +
\end{equation}
\begin{displaymath}
3 x (y + z) ( 5 + 3 y^2 + 3 z^2) + x^2 (10 + 7 y^2 - 5 y z + 7 z^2)
 + y^2 (-11 +14 z^2))
\end{displaymath}
and the  $(1,2)$ off-diagonal entry is
\begin{equation}
-5 x^4 + 14 x^3 y + 2 x^2 (5 + 9 y^2 + 14 y z- 5 z^2) -
5 (-1+y^2+z^2)^2 + 14 x y (-3 + (y +z)^2).
\end{equation}
The remaining cells are obtainable by simple symmetry arguments (for example, 
the (2,2) cell can be gotten by interchanging $x$ and $y$ in (\ref{fds})).
\subsubsection{$N=6$}
For $N=6$, we used an optimal (but not minimal) set of thirty-three
measurements. We found --- using a large number of randomly generated 
points $(x,y,z)$ --- that the associated Fisher information matrix
was strictly dominated by $5 H_{q}(x,y,z)$, but not by $4.99  H_{q}(x,y,z)$.
The Fisher information matrix takes the form (cf. (\ref{cue})) 
\begin{equation} \label{owd}
5 H_{q} (x,y,z) + {1 \over 120} \pmatrix{a & A  x y  &  A x z  \cr
A x y  & b&   A y z  \cr  A x z  &  A y z  & c \cr},
\end{equation}
where 
\begin{equation}
A=193 - 31 r^2, \quad a = - 125 -146 y^2 - 146 z^2 + 31 (y^2 +z^2)^2 
+x^2 (47 +31 y^2 +31 z^2),
\end{equation}
and the diagonal entry 
$b$ can be obtained from $a$ by interchanging $x$ and $y$,  
and $c$ from $a$ by interchanging $x$ and $z$. 

One of the three negative
eigenvalues of the second (``residual'') matrix in (\ref{owd}) is
$(125 -172 r^2 + 47 r^4)/(120 (-1+r^2))$. Now, if we were to rewrite 
(\ref{owd}) in the form of 
$4.99  H_{q}(x,y,z)$ plus a {\it slightly} revised residual matrix,
the eigenvalue in question would be altered only in the respect that 
the constant 125 would change to 123.8. This would render
it {\it positive} for
$r >.992348$, leading to a loss of strict dominance for $r \in 
[.992348,1]$.
 In this specific sense, the upper bound of $5 H_{q}(x,y,z)$
on the Fisher information matrix is {\it tight}.
The residual matrix for $N=4$ strictly dominates that for $N=6$. This 
indicates that the ``fit'' of $(N-1) H_{q}(x,y,z)$ to the Fisher information
matrix for optimal measurements of $N$ copies {\it improves} as $N$
increases.
\subsubsection{$N=7$} \label{ssecn7}
For $N=7$, employing a 42-vector of probabilities, we found the Fisher
information matrix to be strictly dominated by $6 H_{q}(x,y,z)$, but {\it not}
 by
$5.99 H_{q}(x,y,z)$.
Reviewing our previous analyses, we then found that the
analogous situation held also  for $N=3,\ldots,6$, 
that is, the Fisher information
matrix was dominated by $(N-1) H_{q}(x,y,z)$, but not by $(N-1.01)
H_{q}(x,y,z)$. The violations of these 
{\it diminished} bounds occur for nearly pure states, that is 
$r \approx 1$.

 Pursuing this line of thought, 
if we restrict consideration to the more mixed states for which
 $r < {1 \over 2}$, then for $N=7$ we have found that $3.9 H_{q}(x,y,z)$,
but not $3.85 H_{q}(x,y,z)$
bounds the Fisher information matrix for the optimal set of measurements.
Calculations suggest the hypothesis that in the neighborhood of 
the fully mixed
state $r=0$, the bound 
on the Fisher information matrices approaches from above
 $N  H_{q}(0,0,0)/2$, that is 
${N \over 2}$ times the $3 \times 3$ identity matrix. 
Now, the fully mixed state is classical (binomial) in character, while the
pure states are quantum in nature. (It is interesting to note that Frieden
finds that in classical scenarios, only {\it one-half} of the bound or 
phenomenological information $J$ is utilized in the intrinsic 
quantum information $I$ \cite[eqs. (5.39), (6.55)]{frieden}.
``In all covariant quantum theories (e. g., quantum mechanics, quantum 
gravity) $I$ and $J$ are exactly equal. In deterministic classical theories
such as classical electromagnetics and general relativity $I=J/2$. 
But in statistical classical theories $I=J$ again'' [e-mail message
from Frieden].)
\subsubsection{$N>7$}
We are not able to
proceed any further, that is for $N>7$, as there presently do not appear to be
corresponding 
sets of optimal measurements. As a {\it caveat} to the reader, 
let us point out that to recreate the optimal measurements for the
cases $N=6$ and 7 (which unlike the instances $N<6$, were not
formally demonstrated to be minimal in character), 
it is necessary to rely upon the quant-ph preprint version 
(9803066) of \cite{vidal2}, since there are certain errors (as confirmed
in an e-mail from R. Tarrach, though no formal {\it erratum} has 
appeared) in the final, published paper.
\subsection{Properties of the Computed Fisher Information Matrices}
\subsubsection{{\it Diagonal} nature for {\it even} $N$
in spherical
coordinates} \label{dnfe}
We have found that the Fisher information matrices given above 
for the optimal measuements of Vidal {\it 
et al} \cite{vidal} for both $N=4$ and 6
are {\it diagonal} in spherical coordinates ($r,\theta,\phi$).
For $N=4$, this is
\begin{equation} \label{diagn=4}
{1 \over 12} \pmatrix{ {29 + 7 r^2  \over 1 - r^2} & 0 & 0 \cr
0 &  r^2 (29 -  5 r^2) & 0 \cr
0 & 0 & r^2 (29  - 5  r^2) \sin^{2}{\theta} \cr},
\end{equation}
and for $N=6$,
\begin{equation} \label{diagn=6}
{1 \over 120} \pmatrix{ {475 + 172 r^2 - 47 r^4 \over 1 - r^2} & 0 & 0 \cr
0 & r^2 (475 - 146 r^2 + 31 r^4) & 0 \cr
0 & 0 & r^2 (475 -146 r^2 + 31 r^4) \sin^{2}{\theta} \cr}.
\end{equation}
For $N=2$, we also have a corresponding diagonal matrix, that is,
(\ref{sPh}). 

Cox and Reid \cite[p. 2]{cox} have listed three ``consequences of
orthogonality'' of the parameterization of a Fisher information matrix, such
as we have just observed.
These are that: 
(i) the maximum likelihood estimates of the means of the parameters
are asymptotically independent; (ii) the asymptotic standard error for
estimating one parameter is the same whether the other parameters are treated
as known and unknown; and (iii) there may be simplifications in the numerical
determination of the means of the parameters.
``While orthogonality can always be achieved locally, global orthogonality
is possible only in special cases'' \cite[p. 2]{cox}.
In accompanying discussions to \cite{cox}, Sweeting identifies four
advantages to orthogonalization --- computation, approximation, interpretation,
and elimination of nuisance parameters --- while Barndorff-Nielsen, as well as
Moolgavkar and Prentice, 
explain parameter orthogonality in terms of Frobenius' Theorem. The latter
authors also 
indicate that the theorem of de Rham \cite[p. 187]{kobayashi}  gives 
necessary and sufficient conditions for each 
orthogonal parameter to be independent of
the others (as they are {\it not} in our  three even-dimensional examples just 
given).

\subsubsection{Pure- and fully mixed state limits}
Again using spherical coordinates, it is interesting to note
that for the {\it odd} cases of $N=3,5,7$, in the pure state limit
($r \rightarrow 1$), the off-diagonal elements of the corresponding
$3 \times 3$ Fisher information matrix converge to zero. 
In all six (both odd and even) cases, in this same limit, the (1,1)-entries
are  indeterminate, the (2,2)-entries are  ${N \over 2}$ and the 
(3,3)-entries are ${N \sin^{2}{\theta} \over 2}$.

For the fully mixed state, $r=0$
(allowing the angular variables $\theta$ and $\phi$ to remain free), 
the only non-zero entry is the (1,1)-cell.
For $N=2$ it is 1, for $N=3$ it is
\begin{equation}
{1 \over 6} \lgroup 10 + \sin{2 \theta} (\cos{\phi} +\sin{\phi}) +\sin^{2}{\theta}
\sin{2 \phi} \rgroup,
\end{equation}
for $N=4$ it is ${29 \over 12}$, for $N=5$, it is ${(103 + 5 \cos{2 \phi}) 
\over 32}$, for $N=6$ it is ${95 \over 24}$, and for $N=7$,
\begin{equation}
{1 \over 96} \lgroup 456 \cos^{2}{\theta} +7 \sin{2 \theta}
(\cos{\phi} +\sin{\phi}) + \sin^{2}{\theta} (456 + 7 \sin{2 \phi}) \rgroup.
\end{equation}
\subsubsection{Integrals over Bloch sphere of volume elements} \label{volel}
For $N=2$, the integral of the volume element of the 
Fisher information matrix (that is, the square root of the determinant) 
over the (Bloch sphere of)
two-level quantum systems is $\pi^2 \approx 9.8696$, for $N=3$ it is
21.0235,
for $N=4$, it is 
\begin{equation}
{1 \over 441} \sqrt{{29 \over 3}} \pi \lgroup 4705 E(-{7 \over 29}) 
-4194 K(-{7 \over 29}) \rgroup \approx 35.0281
\end{equation}
(where $E$ and  $K$ denote the corresponding elliptic integrals),
for $N=5$, it is 51.0763,
 for $N=6$, it is 69.1253, and for $N=7$, 88.8621.
These 
particular results would be needed for the application to the 
optimal measurements of Vidal {\it et al} \cite{vidal} 
of the universal coding
theorem of Clarke and Barron \cite{cb1}, discussed 
below in sec.~\ref{cuc}.

\subsection{Gill-Massar Traces} \label{relations}
Let us first observe that Gill and Massar \cite[eq.(26)]{gill} 
asserted that the upper (quantum [Helstrom] Cram\'er-Rao) 
bound $N H_{q}$, was {\it not}, in general, 
achievable in a multiparameter setting. This does appear to be 
strictly the case.
 However, our results for $N=2,\ldots,7$ for the
three-parameter $2 \times 2$ density matrices, indicate that --- using the
optimal measurements of Vidal {\it et al} \cite{vidal} --- one can,
by choosing $N$ large enough, come indefinitely close for 
the nearly pure states to this
bound.

To further relate to these analyses of Gill and Massar, we have computed for
$N=2,\ldots,7$, the traces of the product of $H_{q}(x,y,z)^{-1}$, given
in (\ref{inv}), and the Fisher information matrices 
we have obtained using the optimal
measurements of Vidal {\it et al}. (The traces of Fisher information matrices
play a central role in the work of Frieden on the fundamental 
equations of physics \cite[sec. 2.3.2]{frieden}.) 
For the estimation 
of  pure states, Theorem I in \cite{gill}
asserts that this trace quantity 
is bounded above by $N$, while Theorem II there says
that the same bound applies to  mixed states, with the restriction
to {\it separable} measurements. It is also
demonstrated there that these bounds are attainable --- and for large $N$ 
{\it simultaneously} for {\it all} states.

\begin{figure}
\centerline{\psfig{figure=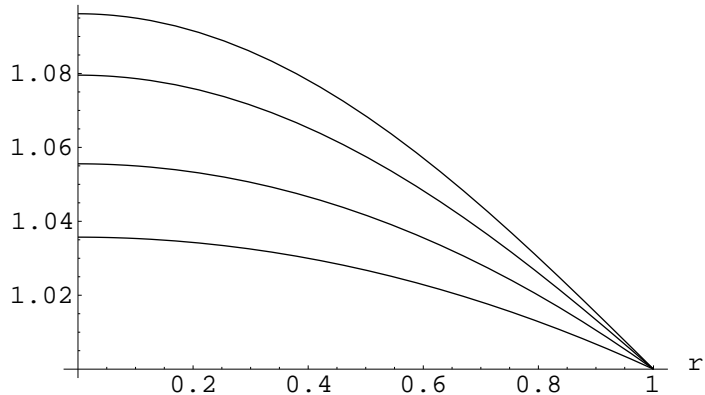}}
\caption{Gill-Massar traces for $N=4,5,6$ and 7 
scaled by their values at the pure states, $r=1$, that is, 
$2 N -1 $. The $y$-intercepts for $r=0$, corresponding to the 
fully mixed state, increase with $N$.}
\label{gmt}
\end{figure}

For $N=2$, it is easy to see, in the context of the 
results above, that this 
(``Gill-Massar'') trace result is simply 3. For $N=3$, we get another 
constant, 5, for the trace.
For $N=4$, we obtain
\begin{equation} 
GM_{4} = {29 - r^2 \over 4},
\end{equation}
 which is 7 for pure states
and 7.25 for the fully mixed state.
For $N=5$, the Gill-Massar trace is
\begin{equation}
GM_{5} = {19 - r^2 \over 2}, 
\end{equation}
which is 9 for pure states and 9.5 for the fully 
mixed state. For $N=6$, it is
\begin{equation}
GM_{6} = {95 - 8 r^2 + r^4 \over 8}.
\end{equation}
This last 
expression is monotonically decreasing from ${95 \over 8} =11.875$ at $r=0$ to
11, that is, $2 N -1$ at $r=1$. For $N=7$, the Gill-Massar trace is
\begin{equation}
 GM_{7} = {57 - 6 r^2 + r^4 \over 4}, 
\end{equation} 
which 
equals ${57 \over 4} = 14.25$ at $r=0$ and 13 at $r=1$, being again
$2 N - 1$.
(In an earlier version of this paper, quant-ph/0002063, the results 
given --- including Fig. 1, plotting the Gill-Massar trace --- for
$N=7$ were ``anomalous'', in this regard. We subsequently ascertained
 that they were erroneous
in nature, due to a programming error.) In Fig.~\ref{gmt}, we plot 
${GM_{N} \over (2 N-1)}$ for $N=4,5,6$ and 7.

It is easy to see, then, that
in these six cases the Gill-Massar bound \cite[eq. (27)]{gill}
of $N$ is violated --- as Theorem III of their paper recognizes will occur
for {\it non-separable} measurements.
So, we obtain a simple pattern of $2 N -1$ for the 
minimum of the trace quantity
in question. 
In regards to these results, R. Gill remarked in an e-mail message 
of Feb. 18, 2000 that
``this is all very interesting. It means that there is a big discontinuity
at the surface of the Bloch sphere (where none of these $3 \times 3$ Fisher
information matrices is well-defined), and it means that the gain in using
joint measurements over separate measurements for mixed states is substantial
throughout the Bloch sphere''.
\subsection{Analyses for $m$-Level {\it Pure} States} 
\subsubsection{$m=2$}
In a further effort to relate to the analyses of Gill and Massar
\cite{gill}, let us consider for the moment simply the two-level pure states,
so we set $r=1$. In terms of the polar coordinates $(\theta,\phi)$,
the Helstrom information matrix takes the  form (cf. (\ref{sPh}), 
\cite[p. 4238]{FUJI})
\begin{equation} \label{fgn}
\pmatrix{1 & 0 \cr
0 & \sin^{2}{\theta} \cr}.
\end{equation}
Then, the Fisher information matrix for the optimal measurements of 
$N$ copies \cite{vidal2} is simply ${N \over 2}$ times (\ref{fgn}), as we 
have confirmed through computations for $N=2,\ldots,7$ (cf. \cite{gill}).
(So, in the pure state case, unlike the mixed state one, 
the quantum Cram\'er-Rao bound of $N$ times 
(\ref{fgn}) is not asymptotically 
approached --- though the Gill-Massar trace bound of $N$ is achievable.)
\subsubsection{$m=3$} \label{tl}
We have also verfied that the same basic additive 
relation holds in the case of the {\it three}-level pure states for $N=2$,
using the formulas in \cite{acin}. Let us use the parameterization of
these states
 in terms
of {\it four} angular variables ($\theta,\phi,\chi_{1},\chi_{2})$ 
employed in \cite[eq. (2.1)]{cavesv},
\begin{equation} \label{spin1param}
| \psi \rangle = \mbox{e}^{\mbox{i} \chi_{1}} \sin{\theta} \cos{\phi}
|1 \rangle + \mbox{e}^{\mbox{i} \chi_{2}} \sin{\theta} \sin{\phi}
|2 \rangle + \cos{\theta} |3 \rangle.
\end{equation} 
 Then, the Helstrom information matrix 
is
\begin{equation} \label{kvi}
\pmatrix{4 & 0 & 0 & 0 \cr
0 & 4 \sin^{2}{\theta} & 0 & 0 \cr
0 & 0 & a & -\sin^{4}{\theta} \sin^{2}{2 \phi} \cr
0 & 0 & -\sin^{4}{\theta} \sin^{2}{2 \phi} & b \cr},
\end{equation}
where (cf. \cite{slatprep})
\begin{equation}
a =  {1 \over 2} \lgroup 6 + 2 \cos{2 \theta} +
\cos{2 (\theta - \phi)} -2 \cos{2 \phi} +
 \cos{2(\theta+\phi)} \rgroup  \sin^{2}{\theta}  \cos^{2}{\phi},
\end{equation}
\begin{displaymath}
b=-{1 \over 2} \lgroup -6 - 2 \cos{\theta} + \cos{2 (\theta -  \phi)}
-2 \cos{2 \phi} +\cos{2(\theta +\phi)} \rgroup
 \sin^{2}{\theta} \sin^{2}{\phi}.
\end{displaymath}
(Note that (\ref{kvi}) is free of the variables, $\chi_{1}$ and
 $\chi_{2}$ --- as (\ref{sPh}) is free of $\phi$.)
So, for $N=2$ copies of a spin-1 system, the Fisher information matrix is
identically (\ref{kvi}), paralleling the specific results for both the pure
and mixed two-level quantum systems for $N=2$. We also intend to analyze 
the case $N=3$, using the specific prescription for the corresponding 
optimal measurements in \cite[sec. 6]{acin}. 
\subsubsection{supplementary analysis for 3-level 
{\it mixed} states} \label{uue}
We have attempted --- following the 
general methodology laid out by Vidal {\it et al} \cite{vidal} 
for the {\it two}-level mixed quantum systems --- to construct
an optimal measurement scheme for $N=2$ copies of mixed 
{\it three}-level 
systems. In doing so, we incorporated 
the optimal measurements for $N=2$ copies of
{\it pure} three-level quantum systems presented 
by Ac\'in, Latorre and Pascual in \cite[sec. 5]{acin}, that
were utilized immediately above. (J. Latorre informs me that he and his
co-authors  ``did not find any manageable way to make progress'' in such 
extended $m=3$ 
{\it mixed} cases, although he did point out that Arvind had recast and 
further developed many of their results using Penrose rays --- in 
apparently yet unpublished work.)
This led us to an oprom with {\it twelve} distinct outcomes, {\it nine}
 corresponding
to the vectors explicitly presented in \cite[eqs. (39), (40)]{acin}, 
and the additional {\it three}
coming from our own orthogonal decomposition of the 
associated rank three 
``residual'' projector
(cf. \cite[eq. (3.3)]{vidal}). (A weight of ${2 \over 3}$ was applied
to the subset of nine outcomes.)

With this twelve-outcome oprom in hand, we found by {\it numerical} means
that the Gill-Massar trace
equalled a constant, 6 (while for $N=2$ 
copies of {\it two}-level systems this trace quantity 
was found in sec.~\ref{relations} also to be a constant, 3). 
(In \cite{slatprep}, we have 
been investigating the possibility of {\it symbolically}
inverting the $8 \times 8$ Helstrom information matrix --- making use of 
a recently-developed Euler angle parameterization of the $3 \times 3$ 
density matrices \cite{byrdslater}. The Gill-Massar trace would, of course, 
be the
trace of the product of this inverse matrix and the Fisher information 
matrix associated with the twelve-outcome oprom.)
This result and our earlier
ones for $m=2$, $N =2,\ldots,7$, 
lead us to conjecture that for non-separable optimal 
measurements of $N$ $m$-level 
quantum systems, the Gill-Massar trace for all $m$ and $N$ is exactly
$(2 N-1) (m-1)$ in the pure state limit, and no less than this for any
mixed state.

Now, for any measurement of a strictly 
pure state itself, the Gill-Massar trace can not exceed $N(m-1)$ by
Theorem I of \cite{gill}. 
(This bound is known to be achieveable for $m=2$ by Theorem  VII 
of \cite{gill}, and for mixed states using separable measurements by
Theorem VI.) So there is a clear discontinuity displayed
by {\it non-separable optimal}
 measurements {\it near} the pure state boundary, as 
well as considerable increased efficiency in estimating strictly mixed or 
impure states through the use of such measurements.
\subsubsection{$m=4$} \label{fl}
We 
have ascertained
the Helstrom information matrix for pure states of {\it four}-level
systems, making use
of the  appropriate analogue of the parameterization (\ref{spin1param})
presented in \cite[eq. (13)]{venki}. The 
six parameters naturally divide into two sets of three, and once again the
entries of the Helstrom 
information matrix are free of the
(three) members of one of the two sets.
\section{Universal Coding} \label{uc}
We can also apply to the three-dimensional family of 
quadrinomial probability distributions
(\ref{qpd}) certain important 
(classical) asymptotic results of Clarke and Barron \cite{cb1}  
pertaining to a number of problems, including those of universal 
data compression  and density estimation. Then,
 we can compare their
formulas with those for the $2 \times 2$ density matrices
(\ref{bloch}), based on the extension to the quantum domain 
of two-level systems by Krattenthaler
and Slater \cite{kratt,kratt2} of this work of Clarke and Barron 
(cf. \cite{jozsa}). (In what follows, we will denote probability distributions
of a general nature by $w$ and more specific ones by $W$, and subscript
 them --- as noted before --- by either $c$
 or $q$ to denote a result stemming from an analysis
 in the classical or 
 quantum domain.)
\subsection{Classical results of Clarke and Barron} \label{cuc}
Clarke and Barron examined the relative entropy 
($N \rightarrow \infty$) 
between a true density
function and a joint (``Bayesian'') density function 
for a sequence of $N$ random variables taken to be the average of the
possible densities (comprising a parameterized family) with respect to a
 (prior) probability 
distribution over this family of density functions. 
The result of Clarke and Barron for the asymptotic relative entropy 
(Kullback-Leibler index) between the true density and the mixture is
\begin{equation} \label{ios}
{d \over 2} \log{{N \over 2 \pi \mbox{e}}} +{1 \over 2} \log{|I_{c}(\alpha)|} -
\log{w_{c} (\alpha)} +o(1),
\end{equation}
where $\alpha$ denotes the $d$-vector of variables parameterizing the
family of 
densities, $w_{c}(\alpha)$ 
 a prior probability distribution used
to average the $N$-fold products of independent identical density functions, 
and  $I_{c}(\alpha)$ the associated $d \times d$ Fisher information matrix.
As applied to our particular
 three-parameter ($d=3$) family of quadrinomial
distributions (\ref{qpd}), with $\alpha  = (r,\theta,\phi)$,
we have
\begin{equation} \label{lcy}
|I_{c}(r,\theta,\phi)| = \lgroup {64 \over 1-r^2}
 \rgroup  r^4 \sin^{2} {\theta}.
\end{equation}
Then, if we choose for the probability distribution, $w_{c}(\alpha)$, 
the particular one
\begin{equation} \label{bpr}
W_{c}(r,\theta,\phi) = \lgroup {1 \over \pi^{2} \sqrt{1-r^2}} \rgroup
  r^2 \sin{\theta}
\quad 
\propto \sqrt{|I_{c}(r,\theta,\phi)|},
\end{equation}
  the asymptotic relative
entropy between the true density and its Bayesian 
(mixture) average assumes the form
\cite[eq. (1.4)]{cb1}
\begin{equation} \label{out1}
{3 \over 2} \log{{N \over 2 \pi \mbox{e}}} +\log{8 \pi^{2}} +  o(1).
\end{equation}
(Let us note that  $r^2 \sin{\theta} \mbox{d} r \mbox{d} \theta 
\mbox{d} \phi$ is  the Jacobian 
determinant of
the transformation from Cartesian to spherical coordinates or, equivalently,
the volume element in spherical coordinates.)
Our particular selection of $W_{c}(r,\theta,\phi)$ 
is   ``Jeffreys' prior'' for this case, that 
is  the normalized (over the Bloch sphere) form
of the volume element ($\sqrt{|I_{c}(r,\theta,\phi)|}$)
 of the Fisher information
metric (cf. sec.~\ref{volel}). 
(The normalization factor, $8 \pi^{2}$, is evident
 in (\ref{out1})). Jeffreys' priors, as shown by Clarke and Barron
\cite{cb1},  fulfill the desideratum of yielding
the common {\it minimax}
 and {\it maximin} of the asymptotic relative entropy.
In the quantum analogue, though, (\ref{bpr}) does not play this 
distinguished role,
although a close (``quasi-Bures'') relative of it does \cite{kratt2,slathall}.
This probability distribution is
\begin{equation} \label{qB}
W_{q}(r,\theta,\phi) = 
.0832258  {\mbox{e} \over 1 -r^2} \lgroup
 {1-r  \over 1 +r} \rgroup^{1 \over 2 r} 
r^2 \sin{\theta}.
\end{equation}
\subsection{Quantum Results of Krattenthaler and Slater for Two-Level Systems} \label{kssec}
Krattenthaler and Slater \cite{kratt,kratt2} have sought to extend the 
general results of Clarke and Barron to the two-level {\it quantum}
systems (\ref{bloch}). They
averaged the $N$-fold 
{\it tensor} products of identical $2 \times 2$ density matrices 
(\ref{bloch}) (rather than averaging the  simple
products  of $N$ {\it random variables}) 
with respect to (spherically-symmetric/unitarily-invariant) 
probability distributions
distributions  of the form $w_{q}(r) r^2 \sin{\theta}$ 
(cf.  \cite[eq. (1.4)]{vidal}).
 The analogue (in terms of the {\it quantum} relative 
[von Neumann] 
entropy) of the Clarke-Barron result (\ref{ios})
is then ($d=3$)
\begin{equation} \label{pew}
 {3 \over 2} \log{ {N \over 2 \pi \mbox{e}}} +
{1 \over 2} \log{I_{q}(r)} -\log{w_{q}(r)} +  o(1),
\end{equation}
where (cf. (\ref{lcy}))
\begin{equation}
I_{q}(r) = {\mbox{e}^2 \over (1 -r^2)^{2}} \lgroup  {1-r \over 1 +r}
 \rgroup^{1 \over r}.
\end{equation}
So, 
\begin{equation}
 I_{q}(r) r^4 \sin^{2}{\theta} = 144.372 W_{q}(r,\theta,\phi)^{2} ,
\end{equation}
which can be compared with its classical counterpart,
\begin{equation}
|I_{c}(r,\theta,\phi)| = 64 \pi^{4} W_{c}(r,\theta,\phi)^2,
\end{equation}
where $64 \pi^{4} \approx 6234.18$.

\begin{figure}
\centerline{\psfig{figure=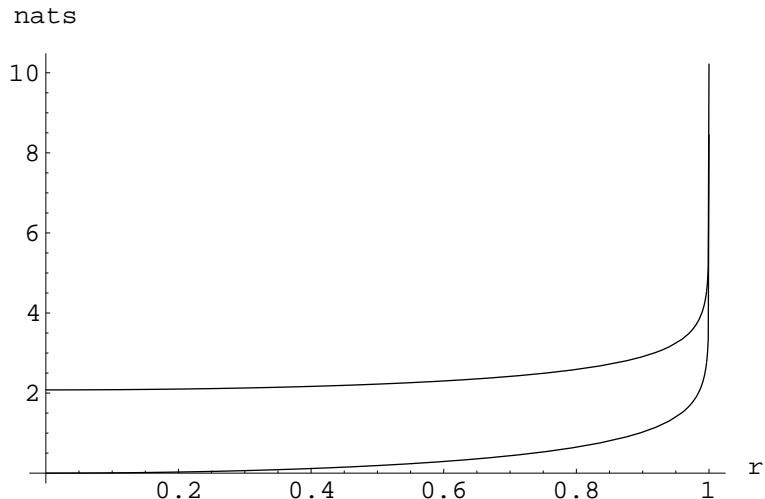}}
\caption{Quantum asymptotic relative entropy 
term --- ${1 \over 2} \log{I_{q}(r)}$ --- and its 
{\it larger} classical 
counterpart, ${1 \over 2} \log{{64 \over 1-r^2}}$, plotted against radial 
distance ($r$) in the Bloch sphere of two-level systems}
\label{nwz}
\end{figure}

As noted \cite{kratt2}, the quasi-Bures probability distribution, $W_{q}
(r,\theta,\phi)$,  given by (\ref{qB}), 
fulfills in the quantum domain of two-level systems
(\ref{bloch}), the distinguished role --- in yielding the common
asymptotic minimax and maximin --- of the Jeffreys' prior (that is, the
volume element of the Fisher information metric) in the classical sector.
In Fig.~\ref{nwz} we plot the term ${1 \over 2} \log{I_{q}(r)}$, 
present in (\ref{pew}), along with the comparable
(but always larger for $r<1$)
classical term, ${1 \over 2} \log{64 \over 1-r^2}$, in (\ref{lcy}). 
The units of the vertical axis are, then,  ``nats'' of information. (A nat
 is equal to $1/ \log_{e}{2} \approx$ 1.4427 bits.) 
So, in the example above, one achieves a lower relative entropy (redundancy)
by proceeding in the quantum domain, as opposed to the classical one.

In the case $r=0$ (the fully mixed state), the 
quantum (Krattenthaler/Slater) asymptotics is given by the expression
\begin{equation}
{3 \over 2} \log{{N \over 2 \pi \mbox{e}}} -\log{w_{q}(0)} + o(1).
\end{equation}
For a pure state ($r=1$), in the case that $w_{q}(r)$
 is {\it continuous} and nonzero
at $r=1$, the asymptotics is given, in general,  by \cite{kratt2}
\begin{equation}
2  \log{N} -3 \log{2} -\log{\pi} -\log{w_{q}(1)} + o(1).
\end{equation}
However, for the particular case of the 
Jeffreys' prior (\ref{bpr}), which is {\it singular} at 
$r =1$, we have \cite[eq. (2.53)]{kratt}
\begin{equation}
{3 \over 2} \log{N} +{1 \over 2} \log{\pi} -2 \log{2}.
\end{equation}

It would be of interest to ascertain if one can construct a
probability distribution for which the (classical) Fisher information 
matrix is equal (in spherical coordinates) to \cite[eq. (3.17)]{petzsudar}
\begin{equation}  \label{bwo}
I_{quasi-Bures} (r,\theta,\phi) 
 = \pmatrix{{1 \over 1 - r^2} & 0 & 0 \cr
0 & {r^2 g(s) \over 1 + r} & 0 \cr
0 & 0 & {r^2 g(s) \sin^{2}{\theta} \over 1 + r}},
\end{equation} 
where $s= {1 -r \over 1 + r}$ and  $g(s) = \mbox{e} s^{{s \over 1 -s}}$. 
(If we employ $g(s) = {2 \over 1 + s}$ in
(\ref{bwo}), we obtain the Helstrom 
information matrix $H_{q}(r,\theta,\phi)$ \cite{petzsudar}.)
This would yield the  {\it quantum} (but non-Helstrom) 
information matrix, the square root of the determinant of which is 
proportional to  the quasi-Bures probability
distribution (\ref{qB}). This probability distribution 
(rather than (\ref{bpr}), as originally conjectured \cite{kratt}) 
has been shown to yield
the common minimax and maximin in the universal coding of the two-level
quantum systems \cite{kratt2}.

\subsection{Relations between {\it Monotone Metrics} and the  
Fisher Information Matrices  Computed in 
Sec.~\ref{omer} } \label{fishmono}
 It would  be of considerable interest to
determine the precise nature $N \rightarrow \infty$
 of the Fisher information matrices 
 corresponding to the use of optimal measurements \cite{vidal}. 
(``For the case of mixed states of spin 1/2 particles, or for higher spins
we do not know what the `outer' boundary of the set of (rescaled) achievable
Fisher information matrices based on arbitrary (non separable) measurements
of $N$ systems looks like. We have some indications about the shape of this
set\ldots and we know that it is convex and compact'' \cite[p. 19]{gill}.) 
In particular, 
we would like to ascertain whether or not there is convergence in 
form (to a diagonal matrix in spherical coordinates) between even and
odd values of $N$, as numerical evidence indicates, 
and whether or not the Fisher information matrices are asymptotically 
simply proportional to some specific 
member (\ref{bwo}) of a broad class of natural 
metric tensors (which includes the Bures and quasi-Bures metrics 
discussed in Sec.~\ref{kssec}) 
for the quantum states associated
with operator monotone functions $f(s) = {1 \over g(s)}$ \cite{petzsudar}.
\subsubsection{The (2,2)- and (3,3)-entries of the diagonal 
Fisher information matrices for even $N$}
\begin{figure}
\centerline{\psfig{figure=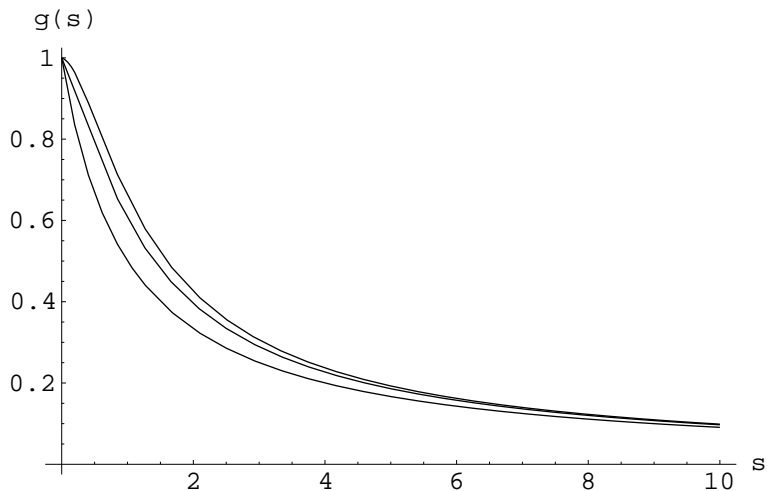}}
\caption{Monotonically-decreasing functions $g(s)$, that is 
(\ref{g(s)4}), 
(\ref{g(s)6}) and (\ref{g(s)2}), obtained by equating
the (2,2)-entries of the computed 
Fisher information matrices (\ref{diagn=4}),  
(\ref{diagn=6}) and (\ref{sPh}) for $N=4,6$ and 2, respectively, with 
$N$ times the 
(2,2)-entry of the general matrix (\ref{bwo}) for a monotone metric.
The curve for $N=6$ dominates that for $N=4$, which in turn 
dominates the hyperbola
for $N=2$.}
\label{gole}
\end{figure}

In fact, if we equate the (2,2)-entries of the diagonal Fisher information
matrices  given in sec.~\ref{dnfe} 
for the optimal measurements for $N=4$ and $N=6$ to the (2,2)-cell
of $N$ times the general matrix (\ref{bwo}) and solve for $g(s)$, 
recalling that $s = {1-r \over 1 +r}$, we obtain for 
$N=4$,
\begin{equation} \label{g(s)4}
g(s) = {1 \over 6 (1+ s)^3}  (6 + 17 s + 6 s^2)
\end{equation}
and for $N=6$,
\begin{equation} \label{g(s)6}
g(s) = {1 \over 45 (1+s)^5} (45 + 222 s + 416 s^2 + 222 s^3 + 45 s^4). 
\end{equation} 
Both these symmetry-exhibiting 
functions, (\ref{g(s)4}) and (\ref{g(s)6}), as well as 
the corresponding 
(Bures/minimal monotone) result (the equation of a hyperbola) 
for $N=2$, that is,
\begin{equation} \label{g(s)2}
g(s) = {1 \over 1 +s}
\end{equation}
 are monotonically-decreasing on the positive real axis 
(Fig.~\ref{gole}), but we are 
presently not aware (for the cases $N=4$ and 6, that is) 
if the reciprocals, $f(s) = 1/g(s)$, are {\it operator}
monotone functions, as required for membership in the class of monotone
metrics of Petz and Sud\'ar \cite{petzsudar} \cite{les}. 
(A function $f(s)$, mapping 
the nonnegative real axis to itself, is called operator monotone if the
relation $0 \leq K \leq H$ implies $0 \leq f(K) \leq f(H)$ for all matrices
$K$ and $H$ of any order. The relation $K \leq H$ implies that all the 
eigenvalues of $H-K$ are nonnegative.)

If we were to include in Fig.~\ref{gole} 
the corresponding function for the {\it quasi-Bures}
monotone metric, that is
\begin{equation}
g(s) = {e s^{s \over 1 -s} \over 2},
\end{equation}
it would be essentially indistinguishable from the hyperbola for $N=2$
(corresponding to the Bures/minimal monotone metric).
\subsubsection{The (1,1)-entries of the diagonal Fisher information matrices 
for even $N$}
\begin{figure}
\centerline{\psfig{figure=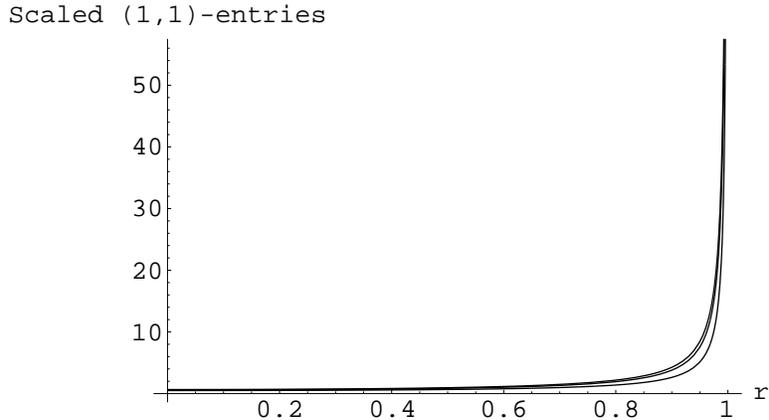}}
\caption{(1,1)-entries divided by $N$ of the 
computed diagonal Fisher information matrices (\ref{sPh}), (\ref{diagn=4}) 
and (\ref{diagn=6}) for $N=2,4$ and 6, respectively. The value at $r=.9$ is 
greatest for $N=6$ and least for $N=2$.}
\label{11entry}
\end{figure}

If, pursuing these lines of thought, one could develop a formula for arbitrary
(even) $N$ for the (2,2)-entry of the Fisher information matrix for optimal
measurements, and 
obviously easily then for the (3,3)-entry (which would be
the (2,2)-entry multiplied by $\sin^{2}{\theta}$), the remaining question, of 
course, 
would be to obtain a general formula for the (1,1)-entry. In this regard,
the apparent general result 
(established above for $N=2,\ldots,7$) 
that the Gill-Massar trace is $2 N -1$ in the 
pure state limit might prove helpful. But since the (1,1)-entry of the metric
tensor for any monotone metric (\ref{bwo}) 
is always simply ${1 \over 1-r^2}$, it would 
apparently be necessary to have some {\it asymptotic} convergence to this 
expression, being that the results 
in the computed Fisher information matrices 
(\ref{diagn=4}) and (\ref{diagn=6}) 
for $N=4$ and 6 (and presumably for
arbitrary even $N$) contain polynomials in $r$ in their numerators, and not
simply a constant term.
In Fig.~\ref{11entry} we plot the (1,1)-entries divided by
$N$ of the computed Fisher
information matrices, in spherical coordinates, for $N=2,4$ and 6.
\subsubsection{{\it Modified} Gill-Massar traces based on the 
Yuen-Lax (maximal monotone) and quasi-Bures  information matrices}
\begin{figure}
\centerline{\psfig{figure=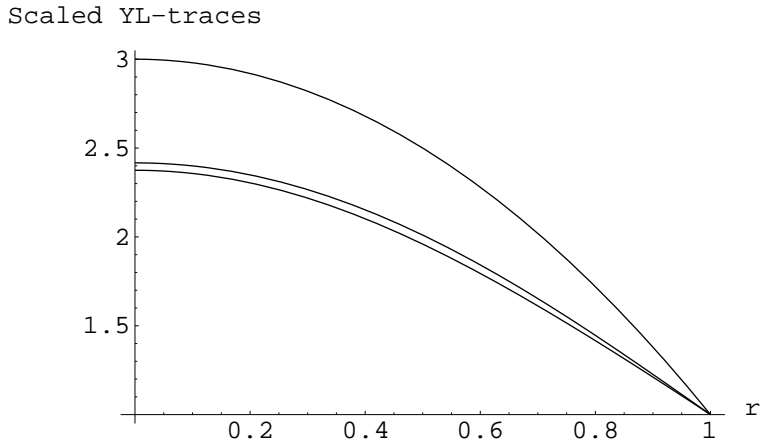}}
\caption{Traces --- scaled by $N - 1$ --- for $N=2,4$ and 6 
based on the Yuen-Lax/maximal monotone metric analysis. 
The $y$-intercepts for $r=0$ 
increase with $N$.}
\label{yleps}
\end{figure}

In sec.~\ref{relations}, we defined the Gill-Massar trace as the trace of the
product of the inverse of the quantum {\it Helstrom} information matrix
and the Fisher information matrices we had  computed 
(sec.~\ref{omer}) based on the optimal
(in terms of {\it fidelity}) measurements of Vidal {\it et al} \cite{vidal} 
for $N=2,\ldots,7$. Now the quantum
Helstrom information matrix corresponds to the use of the {\it minimal}
 monotone
(Bures) metric, as well as the {\it symmetric} logarithmic derivative. Now, 
we replace this with the {\it maximal} monotone metric, corresponding 
to the {\it right} logarithmic derivative \cite[eq. (4.27)]{helstrom}, 
associated with Yuen and Lax \cite{yuen}. 
This can be accomplished by using
$g(s) = {(1+s)/ (2 s)}$ in 
the (diagonal/orthogonal) 
metric tensor (\ref{bwo}) rather than $g(s) = {2 \over 1+t}$ (which gives the
quantum Helstrom information matrix).
Then, we find that in the pure state limit ($r \rightarrow 1$) the values of
the so-modified traces are  exactly $N-1$ --- rather than $2 N - 1$ --- for 
all our six cases
$N=2,\ldots,7$.
For $N=2$, this is
\begin{equation}
\tilde{GM}_{2} = 3 - 2 r^2,
\end{equation}
for $N=4$,
\begin{equation}
\tilde{GM}_{4} = {1 \over 12} (87 - 61 r^2 + 10 r^4),
\end{equation}
and for $N=6$,
\begin{equation}
\tilde{GM}_{6} = {1 \over 120} (1425 -1070 r^2 + 307 r^4 - 62 r^6).
\end{equation}
These three functions, scaled by their value at $r=1$, that is $N-1$, are
plotted in Fig.~\ref{yleps}.
The traces $\tilde{GM}_{N}$ for $N=3$ and 7  are (three-line) 
functions of not only $r$, as previously,
 but of 
$\theta$ and $\phi$ as well. 
For $N=5$, we have 
\begin{equation}
\tilde{GM}_{5} = 
{1 \over 16} ( 147 - 96 r^2 + 13 r^4 + {10 (r^2-1)^3 \over r^2 + r^2 
\cos{2 \theta} -2}).
\end{equation}
In the fully mixed state limit ($r \rightarrow 0$),
the values of the traces are 3, 5, 7.25, 9.5, 11.875 and 11.1875.

\begin{figure}
\centerline{\psfig{figure=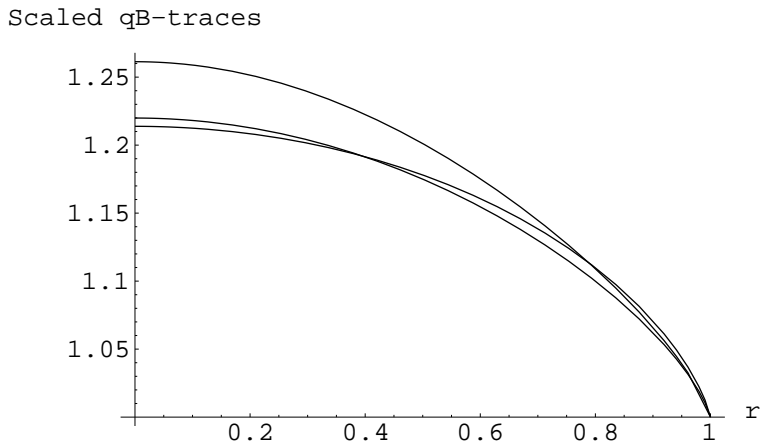}}
\caption{Traces --- scaled by their 
values at $r=1$ --- for $N=2,4$ and 6 
based on the quasi-Bures 
monotone metric analysis. The $y$-intercepts for $r=0$ increase with $N$.}
\label{qbtrace}
\end{figure}

If we alternatively employ the quasi-Bures metric, using
$g(s) = e s^{{s \over 1-s}}$, then, in the pure state limit for 
$N=2,4$ and 6 we get traces equalling $(4 +e)/e \approx 2.47152$, 
$3 + 8/e \approx 5.94304 $ and $5 + 12 / e \approx 9.41455$,
respectively.
(These results are intermediate, then, between those for the minimal 
and maximal monotone metrics.) For $r=0$, the corresponding outcomes
are the same as in the two situations above. In Fig.~\ref{qbtrace}, we plot
these three traces scaled by the noted values at $r=1$.
The curves for $N=2$ and 4 intersect at $r=.395121$.

\section{Concluding Remarks}

We have explicitly constructed the 
$3 \times 3$ Fisher information matrices for the optimal
measurements of Vidal {\it et al} \cite{vidal} for 
$N=2,\ldots,7$, 
found  that they are tightly
bounded by $(N-1) H_{q}$ near the pure state boundary, and
conjectured that they converge from above to 
${N \over 2}$ times the identity matrix at the fully mixed state ($r=0$).
As our main finding, we have uncovered (sec.~\ref{relations}) an 
interesting (less strict)
 analogue for non-separable
measurements of a ``new quantum Cram\'er-Rao inequality'' of 
Gill and Massar \cite[eq. (27)]{gill}. The possibility of extending it to
the cases $N>7$ appears to be a challenging problem.
Also, the development of optimal measurement schemes for multiple copies of
$m$-level systems, $m>2$, and the subsequent evaluation of their Fisher
information characteristics, merits investigation
(cf. \cite{acin}). In this regard, we have presented in sec.~\ref{uue} 
additional evidence --- for an optimal 
measurement we devised for the case $m=3$, $N=2$ --- that 
has led us to the conjecture that for optimal non-separable measurements of 
$N$ copies of $m$-level quantum systems, the ``Gill-Massar trace'' 
equals $(2 N-1) (m-1)$ in the pure state limit for {\it all} $m$ and $N$.

Additionally,
it would be of interest to study the Fisher information matrices associated
with 
optimal measurements based on  {\it continuous} 
oproms \cite[p. 386]{peres} \cite{slaterperes}.
The relation between optimal measurements (sec.~\ref{nce}) and 
universal quantum 
coding (sec.~\ref{kssec})--- both 
involving averaging with respect to isotropic prior
probability distributions by projecting onto total spin 
eigenstates --- appears to be worthy of 
further consideration. (Fischer and Freyberger recently compared 
the use of single adaptive measurements --- which possess certain
practical advantages --- with the use of non-separable ones
\cite{fischer}.)

We have also investigated here several related topics, all pertaining to the
information-theoretic properties of the two-level quantum systems. 
We have posed  the problem of constructing an operator-valued
probability measure (oprom) for 
the smallest number possible of copies $N \geq 4$ 
 which yields the quadrinomial probability
distribution (\ref{qpd}), the Fisher information matrix 
 for which is
simply four times the quantum (Helstrom) information matrix (\ref{qpd}).
Also, we discuss in sec.~\ref{ssecn7}
 what appears to be an intriguing connection between our results
and the work of Frieden \cite{frieden} concerning differences between 
classical and quantum information.

\acknowledgments

I would like to express appreciation to the Institute for Theoretical Physics
for computational support in this research,  as well as 
to M. J. W. Hall, G. Vidal,
R. Tarrach, 
R. Gill and B. R. Frieden for various forms of assistance and advice.

\listoffigures
\end{document}